# The ISGRI CdTe gamma camera
## *In-flight behavior*

François Lebrun

On behalf of the ISGRI team

*Abstract--* **The INTEGRAL Soft Gamma-Ray Imager (ISGRI) is the first large CdTe gamma camera ever built. It provided faultless operations in space since the launch of INTEGRAL in October 2002. A general presentation of the system is given with particular attention to the noisy pixel handling. The observed effect of charge particles on the detectors and their integrated electronics is presented. The in-flight detector evolution is detailed and the camera performance is reviewed.**

## I. INTRODUCTION

IN space, charged particles such as cosmic-ray protons deposit a huge amount of energy in detectors. A very large detector such as the gamma camera of the SIGMA telescope on board GRANAT [1] is crossed several times per millisecond. As a result, the overall performances, and particularly the spatial resolution, are degraded at low energy. Pixel gamma-cameras, where each pixel is an independent detector with its own electronic chain, avoid this problem since the average time between two successive protons in a single detector can be relatively long; allowing for a complete recovery of the electronics. Moreover, the angular resolution of pixel gamma cameras is independent of energy and can be made as good as permitted by the power consumed and dissipated by the large number of electronic chains. This and the need to ensure a lower threshold below 20 keV were the main drivers for the design of new gamma camera for space applications. The spectral performance, the ability to operate at ambient temperature and the technological maturity of the cadmium telluride (CdTe) manufacturing led to the choice of this semi-conductor that was never used to build a large gamma camera neither in space nor even on ground. The preliminary feasibility studies of a large CdTe camera for high energy astrophysics started in 1993 in CEA-Saclay. The illuminating work of Richter and Siffert [2] demonstrating the relationship between charge loss and pulse rise-time has shown that CdTe could be considered as a good spectrometer even for energies above a few hundred keV. However hardware charge-loss compensation based on the pulse rise-time is not applicable to a large CdTe camera housing thousands of detectors with different charge transport properties. On the other hand a precise calibration of the detectors and a pulse rise-time measurement system would allow for a software correction of the charge losses. A laboratory prototype of such a rise-time measuring system was realized and provided excellent results. Then the space aspects had to be considered. There were two concerns. One is related to the possible detector degradation under the effect of charge particle, either cosmic-rays or solar protons. The other is the internal gamma-ray background induced by the spallation reactions in CdTe. Accelerator tests with protons of various energies have shown convincingly that CdTe can sustain relatively high doses provided the dose rate is small, as expected in space [3]. While the contribution of the cosmic diffuse gamma-ray background to the ISGRI count rate was relatively easy to estimate, the internal background contribution was more uncertain. These preliminary studies were encouraging enough to decide the use of CdTe to built a low-energy (20 keV – 1 MeV) gamma camera for astrophysical applications.

The ESA mission INTEGRAL (International Gamma-Ray Astrophysics Laboratory) is a space gamma-ray observatory open to the international scientific community [4]. The satellite payload (see figure 1) is dedicated to the fine spectroscopy (2.5 keV FWHM @ 1 MeV) and fine imaging (angular resolution: 12' FWHM) of celestial gamma-ray sources in the energy range 15 keV to 10 MeV. These performances are provided by the spectrometer SPI [5] and the imaging telescope IBIS [6]. INTEGRAL, with a total launch mass of about 4 t, was launched by a four-stage PROTON from Baikonur/Kazakhstan on October, 17 2002. It was inserted into a highly eccentric orbit (9000 – 154 000 km) in order to provide long periods of uninterrupted observation with nearly constant background and away from trapped radiation (electron and proton radiation belts). Owing to background radiation effects in the high energy detectors, scientific observations are carried out while the satellite is above a nominal altitude of 60 000 km (above the electron belts). The INTEGRAL Science Data Center [7] is in charge of performing the quick look analysis, distribute the

Manuscript received November 12, 2004.
F. Lebrun, ISGRI PI, is with the CEA-Saclay, DAPNIA, Service d'Astrophysique, F91191 Gif sur Yvette Cedex, France (telephone: 33 1 69 08 35 69, e-mail: flebrun @cea.fr).
CEA-Saclay, DAPNIA has realized and maintains in flight ISGRI with the support of CNES.
INTEGRAL is an ESA project with instruments and science data centre funded by ESA member states (especially the PI countries: Denmark, France, Germany, Italy, Switzerland, Spain), Czech Republic and Poland, and with the participation of Russia and the USA.

data and maintain the archive. The INTEGRAL operations are funded till end 2008.

In this paper, we will focus on the in-flight behavior of ISGRI (INTEGRAL Soft Gamma-Ray Imager) [8] the low-energy detection layer of IBIS.

## II. THE IBIS IMAGER AND THE ISGRI DETECTOR

The IBIS detection unit is based on two independent solid state detector arrays, the low energy camera ISGRI (15 keV -1 MeV) and the high energy camera PICsIT (200 keV – 10 MeV) [9]. ISGRI takes advantage of the novel room-temperature semiconductor technology while PICsIT uses scintillators coupled to photodiodes. Both cameras can operate in coincidence to register Compton scattered events.

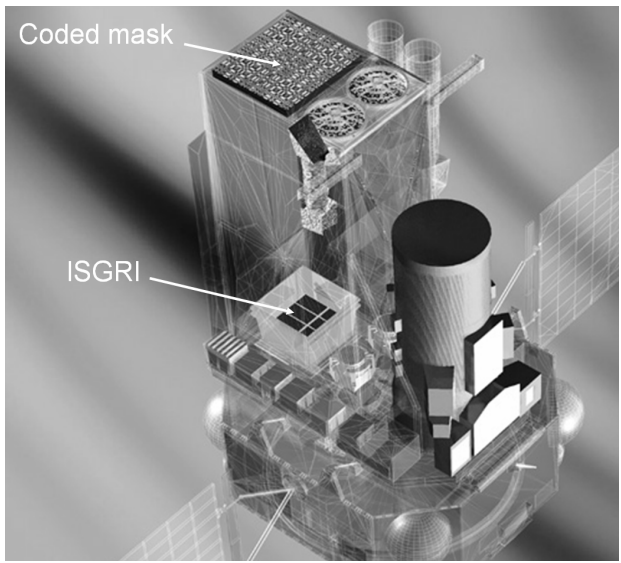

Fig. 1. Schematic view of the INTEGRAL satellite and its payload. The IBIS instrument is basically in two parts: the coded mask and the detection unit. The detection unit comprises two detection layers ISGRI and PICsIT actively shielded by an anticoincidence system. ISGRI is located 3.2 m below the coded mask.

The necessary background reduction is ensured by an active 2 cm thick BGO VETO System surrounding the two cameras. A Tungsten Coded Aperture Mask, 16 mm thick and ~ 1 square meter in area is the imaging device. The angular resolution is ~ 12' FWHM and bright sources can be localized with an accuracy better than 1'. The time resolution is 61 μs. The achieved spectral performance is ~ 9% at 100 keV (ISGRI) and ~ 10% at 1 MeV (PICsIT). We will here focus on ISGRI, the semiconductor camera.

The ISGRI camera is formed with 8 independent Modular Detection Units (MDUs). Each pixel of the camera is a CdTe detector read out by a dedicated integrated electronic channel. Altogether, there are 16 384 detectors (128 x 128) and as many electronic channels. Each detector is a 2 mm thick CdTe:Cl crystal of 4 mm x 4 mm by side with platinum electrodes deposited with an electroless (chemical) process. The ACRORAD company provided 35 000 detectors in total for the various models of ISGRI (breadboard, engineering, qualification, flight and spare). All detectors have been screened for their spectroscopic performance and stability under a 100 V bias at 20 °C. Observed instabilities led to the rejection of about 10% of the detectors. With regard to the spectral performance, the lot has been found very homogeneous. The large difference in the electron and hole mobilities in CdTe implies that the pulse rise-time depends on the interaction depth. A wide pulse rise-time range (0.5–5 μs) is observed and the adopted shaping time (1.5 μs) is a compromise between the response to fast and slow pulses. As a result the ballistic losses are important for slow pulses. In view of the number of independent channels, integrated electronics are necessary. A low noise and low power consumption (2.8 mW/channel) Application Specific Integrated Circuit (ASIC) has been designed for the readout of 4 channels [10]. It allows for the simultaneous measurement of the height and the rise time of every pulse. The RMS noise of the preamplifier is 165 e− (1pF input capacitance) allows good spectral performance down to ~ 12 keV. The layout used a radiation hardened library of components so that the chip is latch-up free and only weakly sensitive to Single Event Upsets. There are many tuning possibilities, the lower threshold (0–70 keV) and the gain of each pixel (± 30%) are both adjustable and each pixel can be disabled. This disabling capability was necessary since a noisy pixel can trigger continuously, resulting in a blind module. A Noisy Pixel Handling System (NPHS) continuously monitors the pixel count rates and can disable or raise the lower threshold of noisy pixels. There is one counter per pixel and one for the whole module. The maximum values attainable by these two types of counters are adjustable. If a pixel counter reaches its maximum value, the pixel is disabled or its lower threshold is raised and all counters are reset. If the module counter reaches its maximum, all counters are reset. On ground, the maximum value for the pixel counter was 3 and the maximum value of the module counter was 100.

## III. ISGRI IN-FLIGHT BEHAVIOR

Previous reviews after 6 [11], 12 [12] and 18 [13] months of operations of the first CdTe camera have already been performed. The present one comes after 24 months of flight. ISGRI is performing very well, and has allowed the detection of more than a hundred sources [14] with major implications on the Galactic emission [15]. The ISGRI images of the sky are the best ever produced at soft gamma-ray energies.

### A. Effect of space particles on ISGRI

However, at the first switch-on of an ISGRI MDU, it was noticed that the NPHS was switching-OFF pixels at a very high rate, nearly one per second, so that the MDU was shortly nearly blind. This was the result of isolated bursts of consecutive triggers in single pixels, appearing randomly in the MDU. These bursts of triggers can be hardly attributed to detector behavior. Their dramatic increase when entering the electron belts proves that belt electrons passing through the mask holes can produce this effect. A study of the ISGRI electronics indicates that they result from preamplifier overload due to large energy deposits (> 10 MeV) attributable either to belt electrons or cosmic-ray protons. Moreover, a recent study shows that these events are able to trigger neighboring pixels and produce peculiar features in a pulse-height – pulse rise-time diagram [16].

The pulse rise-time of the consecutive triggers is always zero. To avoid a telemetry overload, the on-board software was modified to reject events with a pulse rise-time lower than a predefined value (rise-time lower threshold). The maximum number of consecutive triggers observed in a detector was about one hundred. Configuring the NPHS to act on pixels having counted more than 200 while the entire MDU counted less than 10 000 allows for a full recovery of the MDU functionality. However, in this configuration, the NPHS is much less sensitive and the pixel lower thresholds are less accurately adjusted. A fine tuning of the lower thresholds is performed by a software code installed at the ISDC on the basis of the pixel spectra registered in the previous orbit [11]. The few tens of pixels per MDU exhibiting similar but unreasonable spectra were disabled. This pixel behavior (very hard spectrum with no lines) is still not understood. Noisy detectors (excessive counts in the lower channels) are also disabled but are less numerous.

### B. ISGRI In-flight detector evolution

#### 1) Disabled pixels

After an initial period with a fast increase of the number of disabled pixels, some sort of saturation was reached with ~ 400 detectors disabled followed by a regular increase of ~ 1 detector/revolution (see Fig. 2).

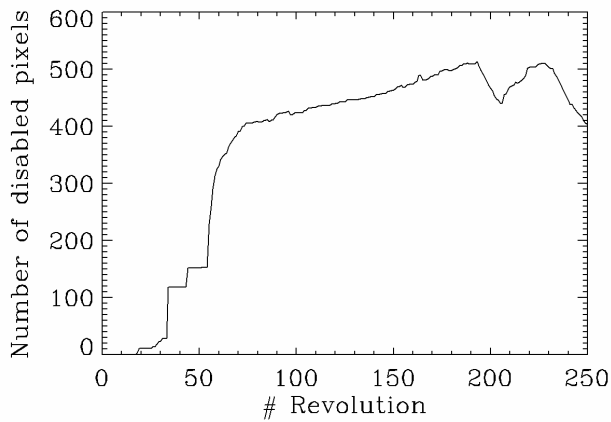

Fig. 2. History of the number of disabled pixels in ISGRI.

After an initial period of manual interventions, a software was deigned to switch off pixels on the basis of their spectral behavior. This software was implemented in revolution 55. A new version of this software allowing to recover previously disabled pixels was implemented in revolution 193. After a month or so, most, but not all, of the recovered pixels, were again automatically switched off. Since revolution 250, the situation is rather stable with ~ 2.5% of disabled pixels.

#### 2) Lower threshold

The lower threshold of the pixels was initially set to step 14 and reached nearly step 15 after the automatic adjustment. After six months, the initial lower threshold was changed to step 13 (revolution 65). A careful ground calibration of the lower threshold was undertaken with the flight operating bias (120 V) and temperature (~ 0°C) and the effective average threshold after the automatic adjustment is given in Fig. 3.

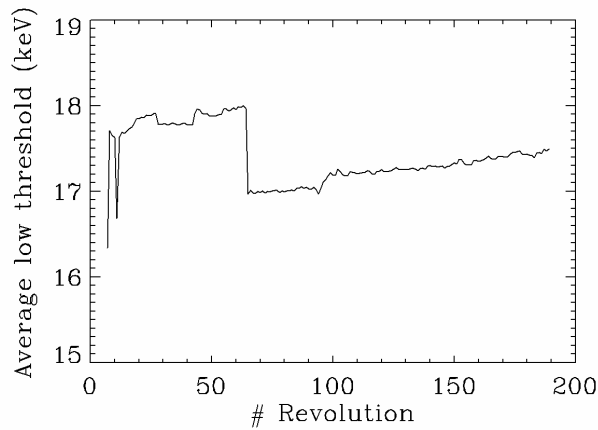

Fig. 3. History of the average pixel lower-threshold.

An accurate knowledge of the lower threshold is essential for the proper reconstruction of the observed source spectra at low energy. The present lower threshold limits the ISGRI sensitivity at low energy. Ground tests in representative flight conditions indicate that it should be possible to decrease the lower threshold by a few keV.

#### 3) Gain evolution

The $^{22}$Na onboard calibration source allows for a continuous monitoring of the spectral performance at 511 keV and at 60 keV (induced W Kα fluorescence line). At high energy, the spectral performance strongly depends on the pulse rise-time, the shorter rise-time events providing the best spectral performance. Using pulse rise-time shorter than 1 µs the in-orbit detector evolution can be first assessed looking at the detector gains. A steady decrease of 2.6 % per year is apparent in figure 4 together with a sharp drop of ~ 0.7 % due to the November 2003 giant solar flare. This observed behavior is fully consistent with the pre-launch accelerator tests using proton irradiation of CdTe detectors [3].

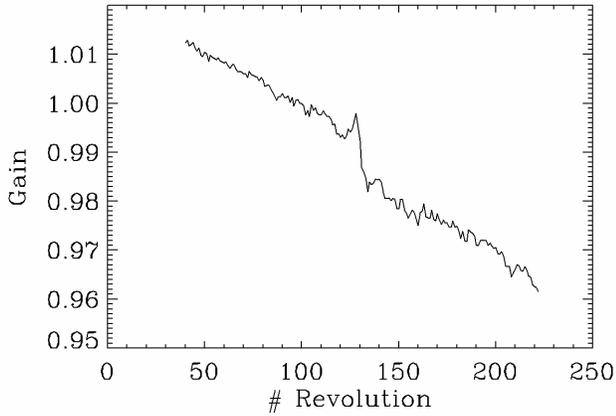

Fig. 4. Gain evolution.

Secondary neutrons may also play a role. The gain loss seems to affect mainly the shorter rise-time events, possibly pointing at an increased electron trapping in the CdTe. There might be also a small offset evolution, by 0.5 keV (see Fig. 5), that is not yet explained nor even fully established.

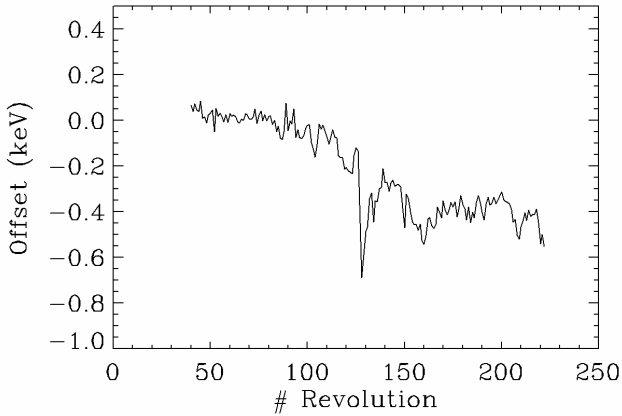

Fig. 5. Possible offset evolution.

Work is on going to understand these results. In any case, the overall behavior of the CdTe confirms that it can be used safely in space with marginal spectral degradation.

### C. ISGRI in flight performance

#### 1) In flight background spectrum

The most critical and uncertain performance parameter was the background spectrum. The low energy part (E < 100 keV) is dominated by extragalactic emission and is relatively well known [17, 18]. On the other hand, the high energy part of the spectrum is dominated by the internal background mainly due to the de-excitation of nuclei produced by spallation reactions of cosmic-rays and solar particles on the instrument. It was uncertain because it could not be extrapolated from a previous CdTe space experiment, since ISGRI is the first one, and because predictions based on Monte-Carlo simulations have not yet reached the required accuracy. It was critical not only because it impacts directly on the experiment sensitivity but also because it could induce a telemetry overflow. Fortunately, this spectrum (see Fig. 6) and the resulting count rates are close to the expectations [3, 20].

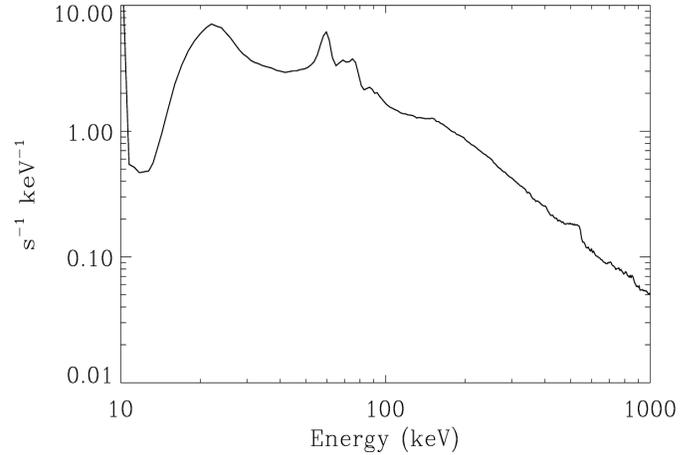

Fig. 6. ISGRI in flight background spectrum..

#### 2) Broad band sensitivity

The sensitivity characterizes the intensity that can be detected by the experiment at a given confidence level (e.g. 3σ) in a given observing time.

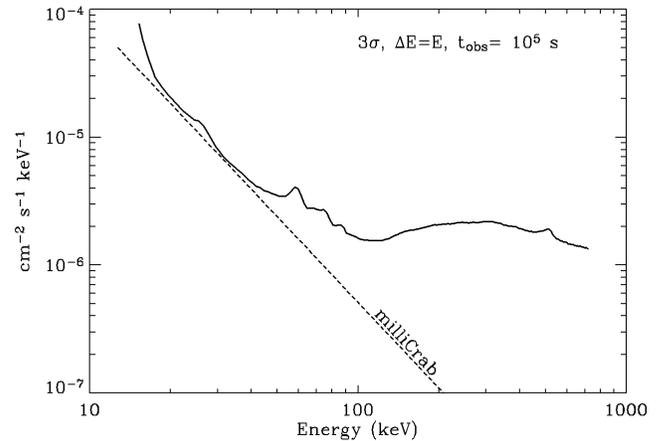

Fig. 7. ISGRI broad-band sensitivity

The sensitivity for continuous spectra is usually defined as the sensitivity in a broad energy band $\Delta E$, e.g. $\Delta E = E$. If systematic errors can be neglected, the sensitivity is limited by the statistical background fluctuations and can be expressed as: $S = n_\sigma (B/ \Delta E/t_{obs})^{1/2} / A_{eff}$. Where $n_\sigma$ is the confidence level, B is the background count rate as given in Fig. 7, $\Delta E$ is the width of the energy band, $t_{obs}$ the observing time and $A_{eff}$ is the effective area. The effective area depends on the detection area, the detector total efficiency, the dead time, the imaging efficiency, the transparency of all material along the photon path and the energy thresholds. The ISGRI statistically limited sensitivity is given in Fig.7. However, for longer observing times, systematic errors may limit the sensitivity. The main source of systematic errors is the residual background structures in the images. Small satellite moves help to disentangle structures due to celestial sources from those due to the background.

*3) Narrow line sensitivity*

The narrow line sensitivity characterizes the intensity of an unresolved line that can be detected by the experiment at a given confidence level (e.g. $3\sigma$) in a given observing time. Its definition is similar to the broad-band sensitivity with two major differences. First, the peak efficiency must be used instead of the total efficiency. Second, it is computed on an energy band equal to the instrument spectral resolution (FWHM).

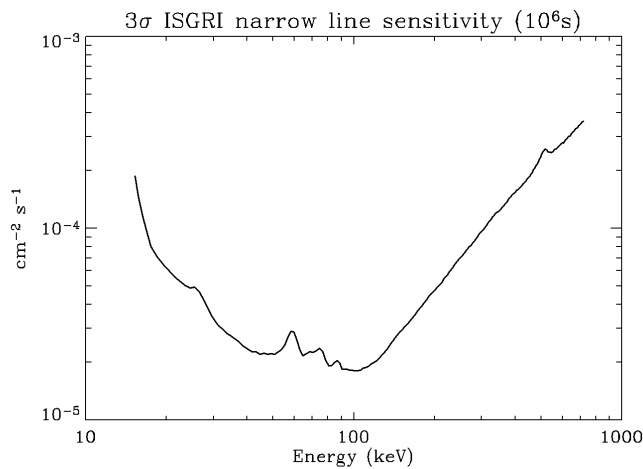

Fig. 8. ISGRI narrow-line sensitivity

### IV. CONCLUSIONS

Today, after 24 months of flight operations, the only sign of CdTe detector degradation is a gain loss of ~2.6 % per year. The 16384 CdTe crystals on board ISGRI are very stable and only ~ 2.5% are disabled because of their noisy behavior. Despite a few unexpected features in the spectral response due to high-energy cosmic-ray particles, the performance of the imager is fully nominal and the background very close to the expectations allows an excellent sensitivity of the telescope in the 15 keV to 1 MeV energy range.

CdTe was known for its very good potential as a gamma-ray spectrometer and it is now proven that a large area detector can be realized and safely used in space. The ISGRI camera produces the best images ever obtained in the soft gamma ray domain. Without any doubt, CdTe and CdZnTe will play in future a key role in instrumental high-energy astrophysics.

### V. ACKNOWLEDGMENT

The author thanks Aymeric Sauvageon for the in-flight data reduction work used to produce the figures of the present paper.